\documentclass[showpacs,superscriptaddress,twocolumn,aps,amsmath,amssymb]{revtex4}

\usepackage{amsfonts}
\usepackage{psfig}
\usepackage{graphics}

\newcommand{\cf}{cf.\ }
\newcommand{\be}{\begin{equation}}
\newcommand{\ee}{\end{equation}}
\newcommand{\bea}{\begin{eqnarray}}
\newcommand{\eea}{\end{eqnarray}}
\newcommand{\Fig}[1]{Fig.\,\ref{#1}}
\newcommand{\Eq}[1]{Eq.\,(\ref{#1})}
\newcommand{\Eqs}[1]{Eqs.\,(\ref{#1})}

\newcommand{\nl}{\nonumber \\}
\newcommand{\eph}{\emph{e-ph}\ }
\newcommand{\mbpar}[1]{\left( #1 \right)}
\newcommand{\mint}[2]{\int\limits_{#1}^{#2}}

\begin{document}
\title{
  Charge-transfer polaron induced
  negative differential resistance and giant magnetoresistance
  in organic spintronics: A Su-Schrieffer-Heeger model study
}

\author{J.~H.~Wei}
\email[Corresponding authors; ]{wjh@sdu.edu.cn, yyan@ust.hk}
\affiliation{Department of Physics, Shandong University, Jinan, China}
\affiliation{Department of Chemistry, Hong Kong University
   of Science and Technology, Kowloon, Hong Kong}

\author{S.~J.~Xie}
\affiliation{Department of Physics, Shandong University, Jinan, China}

\author{L.~M.~Mei}
\affiliation{Department of Physics, Shandong University, Jinan, China}

\author{J.~Berakdar}
\affiliation{Max-Planck Institut f\"{u}r Mikrostrukturphysik,
   06120 Halle, Germany}

\author{YiJing Yan$^{2,\ast}$}

\date{\today}

\begin{abstract}
  Combining the Su-Schrieffer-Heeger
model and the non-equilibrium Green's function
formalism, we investigate the negative
differential resistance
effect in organic spintronics
at low temperature and interprete it with a self-doping
picture. A giant negative magnetoresistance
exceeding 300\% is theoretically predicted
as the results of the negative differential resistance effects.
\end{abstract}


\maketitle

  The discovery of negative differential resistance (NDR)
in traditional semiconductor diodes \cite{Sze90} and also organic
semiconductor (OSE) nanostructures \cite{Che991550,Che001224} has
opened a new chapter of device physics. Motivated by the potential
applications of organic NDR, numerous experiments have been done
with different OSE electronic devices
\cite{Kra022927,Wal041229,Raw03377,Gui0555}. Although some
possible mechanisms have been suggested, the organic NDR remains a
theoretical challenge, beyond the simple picture of interband
tunneling or resonant tunneling in heterostructures. The organic
NDR concerns not only the excess charge transfer through
lead/OSE/lead structures
\cite{Xue997852,Che991550,Kar03165,Emb01125318}, but also the
strong electron-phonon (\emph{e-ph}) coupling that induces
polarons in OSE structures \cite{Xie03125202,Gal05125}.

   Historically, the discovery of magnetoresistance (GMR) effect in 1988 is
considered as the begining of a new technology called spintronics
\cite{Bai882472,Bar908110,Moo953273,Wol011488,Zut04323}, where it
is not only electron charge but electron spin that carries
information. In addition to inorganic semiconductors
\cite{Kik99139,Yan042376}, OSE materials, due to their
controllable structure, strong  \eph coupling \cite{Xie03125202},
and large spin coherence \cite{Kri00173}, offer another promising
system to spintronics \cite{Ded02181,Xio04821}. In a recent
experiment of spin-injection in a spin-valve structure
LSMO/Alq$_3$/Co, a GMR as large as 40\% had been detected
\cite{Xio04821}. In this Letter, we report the possibility of a
charge-transfer polaron-induced NDR mechanism via a numerical
study on a representing $\pi$-conjugated OSE model device that
exhibits also a giant GMR.

  In the study of the electronic conductivity and optical phenomena
in $\pi$ conjugated OSEs, the Su-Schrieffer-Heeger (SSH) model
\cite{Su802099} has also been shown a remarkable track of success. This
model captures the essential characteristic of a conjugated
molecule, where the strong \eph coupling leads it to the
polaron (or solition) charged states and dimerized ground state.
In addition to polyacetylenes, the SSH model has been applied to
charged conjugated systems \cite{Tor00413}, carbon nanotubes
\cite{Wei01207}, and DNA molecules \cite{Con004556, Wei05064304}.
In this work, the SSH Hamiltonian for the OSE electrons coupled
adiabatically with lattice displacements reads
\bea \label{SSH}
  H_O\!\! &=& \!\!
 \sum_{n,\sigma} \Bigl\{
      \epsilon_o c_{n,\sigma}^+c_{n,\sigma}
     - [t_o-(-1)^nt_1-\alpha_o y_n]
\nl &\ & \ \ \times
   (c_{n,\sigma}^+ c_{n+1,\sigma}+{\rm H.c.})\Bigr\}
      +\frac{K_o}{2}\sum_n y_n^2 .
\eea
Each atomic unit in OSE is represented by a single normalized
site; $c_{n,\sigma}^+$ ($c_{n,\sigma}$) denotes the creation
(annihilation) operator of an electron at the $n^{\rm th}$ site
with spin $\sigma$, while $\epsilon_o$, $t_o$, and $t_1$ are the
on-site energy, zero-displacement hopping integral, and
nondegeneracy  parameter, respectively. The lattice distortion is
treated classically, in terms of the bond distances
$\{y_n=u_{n+1}-u_{n}\}$, deviated from its energy-minimum values
that are to be determined via the Hellman-Feynman theorem [\cf
\Eq{HMT}], with the spring constant $K_o$ and the adiabatic \eph
coupling constant $\alpha_o$. This is a static polaron model,
which together with the effective noninteracting many-electron
ansatz in \Eq{SSH}, are considered to be justifiable in the
present study of stationary transport. The effects of polaron
motion and electron-electron correlation (beyond the Hartee-Fock
approximation) will be the subjects of forthcoming work.

  Further, we choose a symmetric ferrimagnetic (FM) $3d$
transition metal as electrodes. The spin-dependent charge
transport takes place between their $3d$ bands. Neglecting
spin-flip during transport and adopting the two-current
model \cite{Fer691784}, we describe the FM metal by one-dimensional
single $d$-band tight-binding model with a spin splitting
term \cite{Xie03125202},
\bea\label{HF}
  H_F
&=&
  \sum_{n,\sigma} \Bigl\{
      \epsilon_f d_{n,\sigma}^+d_{n,\sigma}
       +t_f(d_{n,\sigma}^+ d_{n+1,\sigma}+{\rm H.c.})\Bigr\}
\nl &\ &
  -\sum_n{J_f(d_{n,\uparrow}^+d_{n,\uparrow}
       -d_{n,\downarrow}^+d_{n,\downarrow})},
\eea
where $d_{n,\sigma}^+$ ($d_{n,\sigma}$) is the creation
(annihilation) operator of an electron in the metal at the $n^{\rm
th}$ site with spin $\sigma$; $\epsilon_f$ is the on-site energy
of a metal atom, $t_f$ is the nearest neighbor transfer integral,
and $J_f$ is the Stoner-like exchange integral. The coupling
between the OSE and FM electrodes is described by the
spin-independent hopping integral, $t_{c{\rm L}}=t_{c{\rm
R}}=\beta(t_f+t_o)$, where $\beta$ denotes the OSE-metal binding
parameter.

  The non-equilibrium Green's function (NEGF)
approach based on the Keldysh formalism
\cite{Kel651018,Dat95,Bra02165401} is used to calculate the
quantum transport properties of organic spintronics. To do that,
the spintronic device is divided into three distinct regions. One
is the so-called central scattering region (S-region), with the
Hamiltonian $H_{\rm S}=H_{\rm L}+H_O+H_{\rm R}+H_{\rm int}$, which
consists of the OSE together with a small number of metal atoms
attached to each of its ends. The other two are electrodes (L and
R) that serve as charge reservoirs with the steady state
electronic distribution of bulk metal at given temperature.
Tracing out the reservior degrees of freedom leads to an effective
S-region Green's function $G(E)=[ES-H_{\rm S}-\Sigma_{\rm L}(E)
-\Sigma_{\rm R}(E)]^{-1}$. Here, $S$ (set to be the unit matrix)
is the overlap integral matrix between basis wave functions, while
$\Sigma_{\rm L/R}$ is the self-energy matrix that accounts for the
effects of reservior electrodes on the S-region \cite{Dat95}. It
is possible to have the analytical solution of $\Sigma_{\rm L/R}$
for the one-dimensional FM metal transfer-coupling with the OSE
system. In this work, we adopt an efficient numerical approach
through solving eigenvalue equations to achieve the self-energy
\cite{And918017}. This approach can be easily extended to study
the magneto-transport beyond the one-dimensional system.

  The current can now be evaluated as \cite{Dat95}
\be \label{I-V}
  I=\frac{2e}{h} \mint{-\infty}{\infty}
  {\mbox{Tr} \mbpar{\Gamma_{\rm L} G \Gamma_{R} G^\dagger}}
  [f(E,\mu_{\rm L})-f(E,\mu_{\rm R})] dE.
\ee
The trace term in the integrand
is the transmission coefficient function,
in which $\Gamma_{\rm L/R}=i (\Sigma_{\rm L/R}-\Sigma_{\rm L/R}^\dagger)$
denotes the broadening matrix;
$f(E,\mu_{\rm L/R})$ is the Fermi distribution
function at the lead chemical potential $\mu_{\rm L/R}$.
  One can also evaluate the density of states (DOS) via
$D(E)=i\mbox{Tr}[G(E)-G^\dagger(E)]/(2\pi)$,
and the reduced density matrix as
\be \rho= \frac{1}{2 \pi} \sum_{\alpha={\rm L,R}}
  \mint{-\infty}{\infty}
G \Gamma_{\alpha} G^\dagger f(E,\mu_{\alpha})dE,
\label{DM}
\ee
with a given number of carrier electrons $N=\mbox{Tr}\rho$
in the S-region.
We separate the reduced density matrix
into its equilibrium and bias-induced contributions,
and evaluate them by contour integration and direct
multi-grid Gaussian integration, respectively.
Numerical implementation will be carried out
in a real-function basis set representation,
and the resulting $\rho$ will be real \cite{Bra02165401}.

  We take the bias $V$ not changing the electronic structures of
L and R reservoirs, but just shifting their potentials by $V/2$
and $-V/2$, respectively. In contrast, it does alter the
Hamiltonian of OSE to become $H_O-e\phi(x)$. The electric
potential drop here, satisfying $\phi(x_1)=V/2$ and
$\phi(x_{N_o})=-V/2$, should be evaluated via Poisson's equation
with the method of images \cite{Jac75}
\be\label{Poisson}
  \nabla^2\phi(x)= -\sum_{n=1}^{N_o}
   {\frac{\rho_{n,n}}{\varepsilon_0}\delta(x_n-x)},
\ee
with the charge density $\rho_{n,n}$ that depends also on $\phi(x)$,
the vacuum permittivity $\varepsilon_0$,
and the $n^{\rm th}$ site coordinate $x_n = (n-1)a$, where
$a$ is the OSE lattice constant.

 Furthermore, the OSE lattice distortion, which is correlated with
electron wavefunctions due to the strong \eph interactions, is
assumed to have catched up with the charge variation without
draggling. The resulting lattice distortion at the finite bias
voltage can therefore be evaluated via the Hellman-Feynman
theorem: $\partial [{\rm Tr}(H_{\rm S}\rho)]/\partial u_n =0$;
i.e. [\cf \Eq{SSH}],
\bea\label{HMT}
  2\alpha_o(\rho_{n,n-1}-\rho_{n+1,n})+K_o(y_{n-1}-y_{n}) = 0,
\eea
with the index $n$ including only the OSE sites, since there are
no lattice distortion in the L and R sub-structures. From
\Eqs{DM}--(\ref{HMT}), one can see that both the charging effect
from the electrodes and the external potential from the bias
voltage are all included in the non-equilibrium density matrix of
the coupled \eph system which should be evaluated
self-consistently.

 There are two distinct transport measurement configurations,
parallel (P) and antiparallel (AP), with respect to the relative
magnetization orientation of two FM electrodes. Generally, in the
two-current model, both the majority-majority and
minority-minority (or majority-minority and minority-majority)
transports are permitted in the P (or AP) configuration. But for
the cobalt electrodes in the present work, the full-filled up-spin
electrons cannot transport under in the P-configuration and the
charge carriers are only the half-filled down-spin electrons.
Contrary, the charge carriers in the AP-configuration under $V>0$
are only the up-spin electrons, driven from the full-filled
majority-spin band of L-electrode to the half-filled minority-spin
band of R-electrode of opposite magnitization orientation.

  We are now in the position to
elucidate numerically the $I$-$V$ characteratics,
especially the NDR behavior of OSE systems. The SSH
parameters for the OSE system [\Eq{SSH}] are
$\epsilon_o = -4.3$ eV, $t_o=2.5$ eV,
$t_1=0.04$ eV, $\alpha_o=5.0$ eV/\AA, and $K_o=21.0$ eV/\AA$^2$,
which are the modified values from those of conducting polymers
\cite{Hee88781} for the OSE being of larger band gap.
The tight-binding parameters for the cobalt electrodes
[\Eq{HF}] are
 $\epsilon_f=-7.0$ eV, $t_f=1.5$ eV and $J_f=1.45$ eV,
determined by fitting the cobalt $3d$ band structure \cite{Pap86}.
The OSE-metal binding parameter $\beta$ will be specified later.

\begin{figure}
\includegraphics{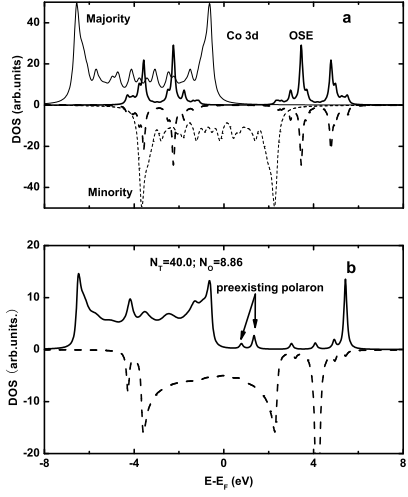}
\caption{ DOS of the majority-spin (solid-curves)
  and minority-spin (dash-curves) bands
   of model systems:
 (a) The isolated OSE (thick-curves)
   and Co $3d$ band (thin-curves),
   with the OSE-metal binding parameter $\beta=0$;
 (b) The binded Co/OSE/Co spintronic device with $\beta=0.5$.
   Temperature $T=11$ K.
}
\label{fig1}
\end{figure}

 To justify the above values,
we calculated the band structures of the model OSE and FM cobalt
electrodes individually (with $\beta=0$); see \Fig{fig1}a for
their resulting DOS at $T=11$ K. Our model OSE of 10 sites has
the highest occupied and lowest unoccupied molecular orbitals
of $E_{\rm HOMO}=E_F-0.9$ eV and $E_{\rm LUMO}=E_F+2.0$ eV,
where $E_F$ is the Fermi level of the cobalt electrode.
These results well reproduce the band gap diagram of the OSE
spin-valve device used in a recent experiment \cite{Xio04821}. The
model cobalt metal is also consistent with the real material
\cite{Pap86}, which shows one full-filled majority-spin
band, and one half-filled minority-spin band.

  Let us start with the equilibrium property of the model
spintronic system at $V=0$, at which the S-region (Co/OSE/Co)
assumes charge neutral. In our model, the S-region consists of a
cobalt oligomer of 10-sites to each terminal of the 10-site OSE;
thus each segment in Co/OSE/Co involves 20 spin-orbitals. If the
OSE-metal binding parameter $\beta=0$, three segments of the
S-region would be charge neutral individually, with $N_{\rm
L}=N_{\rm R}=15$ and $N_{O}=10$ electrons, respectively. In
reality $\beta\neq 0$ and {\it intra-regional charge transfer}
(ICT) is possible. Depicted in \Fig{fig1}b is the calculated DOS
of the organic spintronic device with the OSE-metal binding parameter
$\beta=0.5$, where an ICT of about $1.14e$ from OSE to Co-segments
has occured. The observed `self-doping' phenomenon here is largely
due to the strong \eph interaction, which leads to the
formation of a preexisting hole polaron that stabilizes the
S-region complex before applying potential bias \cite{Bre85309}.
 The preexisting polaron state is rather evident
 by examining the majority-spin
 band of Co/OSE/Co complex (solid-curve in \Fig{fig1}b), since
 its isolated metal counterpart (thick-solid curve in \Fig{fig1}a)
 is completely filled up to the Fermi level.

\begin{figure}
\includegraphics{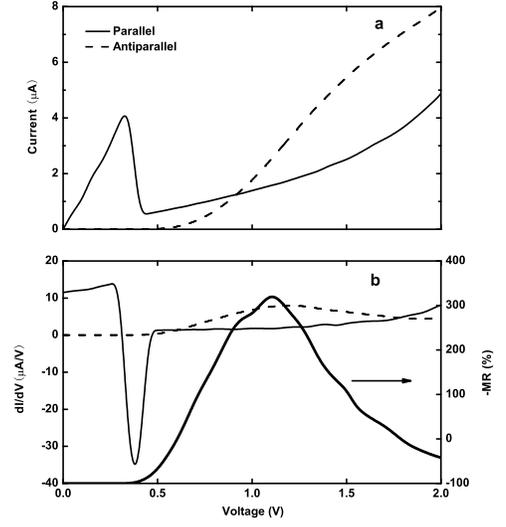}
\caption{(a) The $I$-$V$ characteristics of the Co/OSE/Co system
    in \Fig{fig1}b,
  measured with the P- (solid-curve) and AP-configuration (dash-curve)
  of relative magnetic orientation of electrodes.
  (b) The $dI/dV$, obtained numerically via (a),
     and $-\Delta R/R$ (thick-line) as functions of the bias voltage.
}
\label{fig2}
\end{figure}

  We then calculated the $I$-$V$ characteristic [\Eq{I-V}]
of the model Co/OSE/Co spintronics
at $T=11$ K and $\beta=0.5$,
in both the P and AP configurations
of relative magnitization orientation of FM electrodes.
The resulting $I$-$V$ curves in these two configurations
(P: solid; AP: dash) are shown in \Fig{fig2}a,
and the corresponding $dI/dV$ ones are in \Fig{fig2}b.
Included in \Fig{fig2}b is also a thick-curve
for the bias voltage dependence of
magnetoresistance (MR), $\Delta R/R=
(R_{\rm AP}-R_{\rm P})/R_{\rm AP}$, measuring the relative
difference of electric resistance with these two configurations.

\begin{figure}
\includegraphics{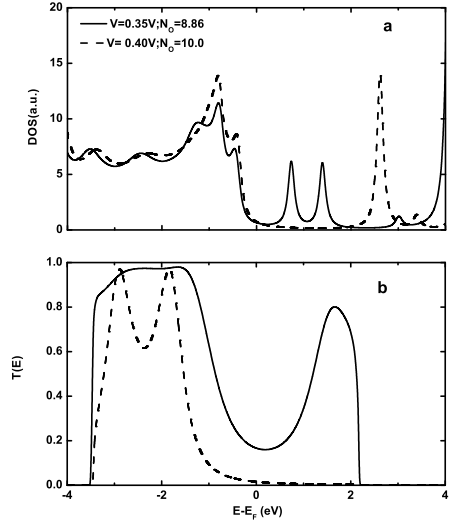}
\caption{The current-peak state (solid-lines) and the valley state
   (dash-lines), referring to the $I$-$V$ charecteristic
    of the parallel configuration in \Fig{fig2}a, in terms
   of (a) the DOS of majority-spin band, and (b) the transmission coefficient of
  electrons (minority-spin).
   The total numbers of  electrons in the OSE
   segment in the current-peak and valley states are specified.
} \label{fig3}
\end{figure}

  Consider first the NDR behavior (about $-26.3$\,\mbox{k$\Omega$} at its
minimum) in the P-configuration,
where the current, after an initial near-ohmic increase,
drops quickly from $I_{\rm peak}=4.45$\,\mbox{$\mu$A} at
$V_{\rm peak}=0.35$\,V to
$I_{\rm valley} = 0.49$\,\mbox{$\mu$A} at $V_{\rm valley}=0.4$\,V.
To see what happens during the NDR region, we also examined other
nonequilibrium properties (at $V>0$). Shown in \Fig{fig3} are the
representing results of both the current peak (solid) and
valley (dash) states: (a) the majority-spin (up-spin) DOS $D(E)$;
(b) the P-configuration (down-spin)
transmission coefficient function $T(E)$.
Indicated in \Fig{fig3} are also the numbers of electrons
in the OSE segment at the two corresponding voltages. By checking
the charge distribution and the lattice distortion (not shown
here), we found: (i) The preexising hole
polaron remains localized around the OSE center
when $0\leq V \leq V_{\rm peak}$;
(ii) As $V$ increases further, the excess electron
charge migrates from the leads into the OSE segment;
and (iii) At $V_{\rm valley}=0.4$ V,
  the preexisting hole is completely annihilated,
 and the OSE is essentially in its dimerized ground charge-neutral state.

  The above observations suggest
that the NDR behaviour in the P-configuration, shown by the solid
curves in \Fig{fig2}, is due to the annihilation of the
preexisting, `self-doping' polaron.
As discussed earlier (cf.\ \Fig{fig1}b), the
preexisting hole polaron level at $V=0$ deeply localizes in the
down-spin band (P-configuration conduction band) gap of the OSE,
which leads to its relatively large DOS, and
thus a low-resistance state according to
the doping theory of conducting polymers \cite{Hee88781}. This
accounts for the rapid increase of current when $V<V_{\rm peak}$ in
the P-configuration (the solid curve in \Fig{fig2}a).
Figure \ref{fig3} shows clearly that the increase of bias voltage
from $V_{\rm peak}$ to $V_{\rm valley}$ accompanies with the
annihilation of the preexisting polaron, which  accounts for the
NDR observed in the P-configuration (solid curves) in \Fig{fig2}.

  In the AP configuration, the charge carriers at $V>0$ are no
longer the down-spin electrons, but the up-spin ones,
from the majority subband of L  to the minority subband of R electrode.
As the observed switch-on voltage in this case (about 0.5\,V)
exceeds the aforementioned NDR region,
the `preexisting polaron' makes no
direct contribution to the conductance in the AP configuration.
The resulting $I$-$V$ characteristic (dash-curve in \Fig{fig2}a)
can thus be all understood (including its switch-on voltage)
by examining the structures of the two involving
subbands in \Fig{fig1}b at $V=0$.

  Finally, let us make comments on the voltage-dependent MR
(the thick-curve in \Fig{fig2}b), especially the negative GMR of
$(-\Delta R/R)_{\rm max}$ = 300\% at $V=1.1$\,V. Negative MR has
been experimentally observed in the Co/SrTiO$_3$/LSMO tunnel
junction \cite{Ter994288} and the LSMO/OSE/Co spin-valve device
\cite{Xio04821}. Traditionally, one analyses the observed MR, or
other transport behavior as function of bias potential, via the
involving DOS of conduction bands/subbands of uncorrelated
($\beta=0$) FM/OSE/FM systems at $V=0$, such as \Fig{fig1}a and
Ref.\ {\onlinecite{Pap86}}. This traditional analysis does lead to
our understanding the qualitative MR-$V$ behavior in each
individual voltage range in \Fig{fig2}b, where $\Delta R/R$
decreases from 100\%, changes sign into the negative MR region,
reaches $(-\Delta R/R)_{\rm max}$ at $V=1.1$\,V, and etc. However,
the quantitive GMR values in \Fig{fig2}b, especially its maximum
value of 300\%, cannot be accounted for via the simple band
structure analysis with $\beta=0$ and/or $V=0$. The calculated
extraordinary GMR of 300\% can only be accounted for via the
`self-doping' of preexisting polaron and its annihilation that
affect distinctly differently on conductances depending on the (P
or AP) relative magnetization orientation of FM electrodes.

  In summary, we proposed to exploit
`self-doping' FM$_1$/OSE/FM$_2$ structures for distinct NDR and
GMR materials. Numerical demonstrations were performed based on a
realistic model system. The theoretical NDR was found as a result
of transition from the low-resistant preexisting hole (cation)
polaron to high-resistent dimerized charge-neutral ground state.
This NDR mechanism is different from the two-step reduction
picture, proposed originally by Reed and Tour and co-workers
\cite{Che991550} in explaining their experiment on a
redox-center-containing molecule, in which NDR results as
transition from conducting anion to insulating dianion state. As
it appears in the P- but not AP-configuration of relative
magnetization orientation of FM electrodes, the NDR leads also to
a large magnitude (300\%) of negative GMR, much larger than the
experimentally reported ones by far \cite{Xio04821}. Many OSEs are
easy-doping materials \cite{Hee88781}, and the required
`preexisting polaron' state can be formed in either self-doping or
external doping manner. Thus, experimental realizations of GMR far
exceeding 100\% in organic spintronic devices should be feasible
according to the present model study.

    Support from the Research Grants Council of  the Hong Kong Government
 (604804)and the National Natural Science
Foundation of China (Grant No.~10474056) are
gratefully acknowledged.

\clearpage

\clearpage

\end{document}